\documentclass{aa}
\usepackage{amsmath}
\usepackage{graphicx,txfonts,natbib,color}
\bibpunct{(}{)}{;}{a}{}{,}

\begin{document}
\title{Stability of multiplanet systems in binaries}
\author{F. Marzari\inst{1}, G. Gallina\inst{1}}
\institute{
  Dipartimento di Fisica, University of Padova, Via Marzolo 8,
  35131 Padova, Italy
}
\titlerunning{Dynamical stability in binaries}
\authorrunning{F. Marzari and G. Gallina}
\abstract 
{When exploring the stability of multiplanet systems in binaries, two
parameters are normally exploited: the critical semimajor axis $a_c$
computed by Holman and Wiegert
(1999) within which planets are stable against the binary perturbations,
and the Hill stability limit
$\Delta$ determining the minimum separation beyond which two planets
will avoid mutual close encounters.   
Both these parameters are derived and used  in different contexts, i.e. $\Delta$ 
is usually adopted for computing the stability limit of two planets
around a single star while $a_c$ is computed
for a single planet in a binary system.}
{Our aim is to test whether these two parameters can be safely applied 
in multiplanet systems in binaries 
or if their
predictions fail for particular binary orbital configurations. 
}
{We have used the frequency map analysis (FMA) to measure the diffusion of 
orbits in the phase space as an indicator of chaotic behaviour. 
}
{First we revisited the reliability of the empirical 
formula computing $a_c$ 
in the case of single planets in binaries and
we find that, in some cases, 
it underestimates by 10--20\% the 
real outer limit of stability and it does not account for planets 
trapped in resonance with the companion star well beyond $a_c$. 
For two--planet
systems, the value of $\Delta$ is close to that computed 
for planets around single stars, but the level of chaoticity close to it 
substantially increases for smaller semimajor axes and higher 
eccentricities of the binary orbit. In these configurations 
$a_c$ also begins to be unreliable and non--linear secular resonances 
with the stellar companion
lead to chaotic behaviour well within $a_c$, even for 
single planet systems. 
For two planet systems, the superposition of mean motion 
resonances, either mutual or with the binary companion, and 
non--linear secular resonances may lead to chaotic behaviour 
in all cases. 
We have developed a parametric semi--empirical formula determining 
the minimum value of the binary semimajor axis, for a given 
eccentricity of the binary orbit, below which stable two 
planet systems cannot exist. }
{The superposition of different resonances between 
two or more planets and the binary companion may 
prevent the existence of stable dynamical configurations 
in binaries. As a consequence, care must be devoted when 
applying the Holman and Wiegert
(1997) criterion and  the Hill stability against mutual close encounters
for a multiplanet system in binaries. 
}


\maketitle

\section{Introduction} 
\label{intro}

According to \cite{horch2014}, an estimated fraction of about 40 to 50\% of 
exoplanets are in binary star systems. The dynamical properties of 
planets in binaries, in terms of eccentricity distribution, 
do not appear significantly different from those of 
planets around single stars.
It is thus expected that planet--planet
scattering is  effective in leading to physical collisions 
between planets, ejection, and eccentricity pumping in binaries 
as in single stars.  \cite{marzari2005} have shown that there 
are some statistical differences in the outcome 
of the chaotic phase which strongly depend on the 
eccentricity of the binary system for any given value of 
the binary semimajor axis.  For example, in a system of 
three planets on initially unstable orbits, the number of 
surviving systems with two planets in a stable configuration at the end 
of the scattering phase declines with the binary eccentricity.  

In this paper we focus on the conditions for the stability/instability 
of  multiplanet systems in binaries. 
In this study of the dynamics of two planets orbiting the primary 
star of a binary system we test whether 
the criterion for the Hill stability, against mutual close approaches,
is affected by the presence of the 
companion star. 
The initial threshold separation $\Delta$ between two planets
around a single star granting their long term dynamical stability 
is given by $\Delta \leq 2 \sqrt 3 R_{Hill}$, 
where 
$\Delta = a_2 - a_1$ is the difference between the semimajor axes of the 
two planets and $R_{Hill}$ is the mutual Hill sphere defined as 

\begin{equation}
R_{Hill} = \left(\frac{m_1 + m_2} {3 M_{\odot}}\right )^{1/3}
\left( \frac{\left(a_1 + a_2\right)} {2}\right)
\label{eq:cha2}
\end{equation}

This criterion was derived from \cite{glad93} from the work 
of \cite{marchal82} for circular and coplanar orbits and it
was numerically confirmed 
by \cite{CH96}.
More complex 
equations are available when the planets are on initially eccentric 
and inclined orbits \citep{donnison06,veras2013}, but we will concentrate in 
this paper on bodies initially on almost circular orbits 
approximately lying on the same plane. 

In this paper we 
investigate the validity of the above mentioned stability condition 
in binary systems using the same numerical approach of 
\cite{mar20142pia} based on the application of the 
frequency map analysis (hereafter FMA).
The FMA method 
\citep{lask93, sine96, marz03} allows us to outline the stability 
regions in the phase space with short term numerical integrations
by computing the diffusion of the main frequencies of the 
system. This allows a massive exploration of the region close 
to the Hill stability separation to test the influence 
of the binary orbital parameters on its 
value. Before testing the stability of 
two planets in binaries we apply the method to the case 
of a single planet as a test bench. In this way we can compare 
the outcome of the FMA exploration with the 
empirical formula of \cite{holman97} which defines a critical
 semimajor axis $a_c$ within which the orbit of a planet
is assumed to be stable against the binary perturbations. 
We then concentrate on systems with two planets and 
test their stability as a function of the binary parameters
such as semimajor axis and eccentricity of the stellar pair and 
their mass ratio.

In Section 2 we briefly summarize the numerical model used 
to explore the stability of planetary systems in binaries. 
In Section 3 we test the method on single planet systems 
in binaries and compare the results with the semi--empirical formula of 
\cite{holman97} and \cite{ines}. In Section 4 we study two--planet systems 
in binaries and examine the validity of the Hill  
criterion for different parameters of the binary. 
In Section 5 we exploit a statistical approach to derive 
a semi--empirical formula predicting the minimum semimajor 
axis of the binary allowing a significant stability 
region for two planets as a function of the binary eccentricity, 
mass ratio and inner planet semimajor axis.
We summarize and
discuss our findings in Section 6. 

\section{The numerical approach}
\label{model}

The FMA is a powerful tool to measure, with short term 
numerical integrations, the diffusion velocity of 
a dynamical system in the phase space \citep{benettin76}. 
It can be successfully used to explore the stability of two planets 
in a binary system since it allows a fine sampling of the phase space
with a limited computer load. 

We apply the FMA to the non--singular variables $h$ and $k$, 
defined as $h = e cos(\varpi)$ and $k = e sin(\varpi)$, of the 
inner planet. The main frequencies present in the signal are 
due to the secular perturbations of the companion star and 
of the second planet. Each dynamical system analysed with the 
FMA  is numerically integrated 
over 10 Myr, with a sampling period of 5 yr. To perform the 
analysis, the whole 
timespan is divided in time--windows extending
for $5 \times 10^5$. Each window is shifted forward in time by 
$1 \times 10^5$ yr respect to the previous one.  On each of these windows 
the main frequency is computed 
with the FMFT high precision algorithm described in \cite{lask93} and 
\cite{sine96}. The chaotic diffusion of the orbit is measured 
as the logarithm of the relative change of the main frequency of the 
signal over all the windows,  i.e. $c_s = log {{\sigma_f}  \over {f}}$ 
where $\sigma_f$ is the standard deviation of the main
frequency $f$.  Small values of $c_s$ imply a low diffusion 
rate and then stability over a long timescale. 
On the contrary, large values are characteristic of 
systems where the secular frequencies change on a short timescale
and are then chaotic. 
We always test that in all our simulated systems, where different 
values of the binary and planetary orbital elements are adopted, 
the secular frequencies complete at least one circulation period over 
each time--window. This is the minimum interval of time for a
reliable determination of the main secular frequency in the 
spectrum of the $h$ and $k$ variables with the FMA. 
The computer code halts if this condition is not met.  
Indeed, in all our 
models the secular frequencies covered many 
circulation periods per window allowing an accurate determination of
the frequency.

The FMA method could also be applied to the second planet as well and 
an additional value of $c_s$ could be derived from the analysis of its frequencies. 
However, we found that the $c_s$ value for the outer planet is 
systematically slightly worse,  possibly due to the fact that 
in the power spectrum additional frequencies due to the binary 
companion are stronger. This leads to a reduction in the strength
of the peak we use for the frequency analysis, leading to a degradation in 
the numerical computation of the frequency. For this reason, 
we prefer to apply the FMA to the inner planet. 

We will consider two different models, the first, used as a test
bench, is composed of a single Jupiter--size planet in orbit around 
a solar type star member of a binary system. The outcome of 
the FMA in this model is compared to the 
predictions of the empirical formula of \cite{holman97}. 
This formula computes the critical semimajor axis $a_c$ of a planet
in a binary system beyond which instability is expected. It is 
the outcome of a fit to the results of numerical integrations 
where massless particles (elliptical restricted three--body problem) 
are integrated 
over 300 binary periods. Those particles which survive till the end of
the integration outline the stability limit and then $a_c$. The 
critical semimajor axis depends on the physical and orbital parameters
of the binary as it follows:

\begin{align}
a_c = & [ ( 0.464 \pm 0.006) + (-0.380 \pm 0.010) \overline{\mu}   \\  \nonumber
      & + (-0.631 \pm 0.034) e_B + (0.586 \pm 0.061) \overline{\mu} e_B  \\  \nonumber
      & + (0.150 \pm 0.041) e_B^2 + (-0.198 \pm 0.074) \overline{\mu} e_B^2 ] a_B, 
\label{eq:HW}
\end{align}

where $\overline{\mu} = m_2 / (m_1 + m_2) $ (here we term mass ratio 
the value of $\mu = m_2/m_1$), $e_B$ and 
$a_B$ are the eccentricity and semimajor axis of the binary, 
respectively. 

We have analysed a set of dynamical configurations with a 
single planet and compared the prediction of the  
empirical formula of \cite{holman97} with the outcome
of the FMA analysis. Some differences are expected since 
not only we integrate the full elliptical three--body problem with a massive 
planet but we also allow the planet to be on an orbit inclined 
respect to that of the binary up to a value of 
$5^o$. 
In addition, with the FMA approach we can examine 
a significantly larger portion of the phase space with a
chance of a better definition of the stability limit. 

Once performed the test with a single planet, 
we will finally focus on systems made of two planets having the 
same size (that of Jupiter) and evolving on progressively more 
separated orbits.  
We fix the semimajor axis 
of the inner planet to $a_1 = 3$ AU and randomly sample 
the orbit of the second planet.
This method is similar to that usually exploited to test the 
stability against close encounters in multiplanet systems
\citep{CH96, MW02, CHATTERJEE08, smith2009, smith2010, 
mar20142pia, morrison2016}. As shown in \cite{morrison2016} the 
problem does not easily scale with the mass ratio between the star 
and the planets. In our case we have the additional problem that 
the fourth body is a star with a mass comparable to the 
primary star, so the dynamical system cannot be 
easily compared to one made of three planets with equal or similar mass. 
For this reason we have to perform an independent 
analysis.

As standard model we consider a binary system with two equal stars 
of solar mass ($\mu = 1$)
and two different values of binary eccentricity, i.e. $e_B = 0$ 
and $e_B = 0.4$. To perform a statistical analysis of the
dependence of the stable region for two planets as a function of
the binary parameters we consider also systems with
different mass ratio ranging from 0.1 to 1, and with sampled
values of both $a_B$ and $e_B$.

\section{Test on single planet systems: evaluation of the critical 
semimajor axis}
\label{1pla}

\begin{figure}[hpt]
  \includegraphics[width=\hsize]{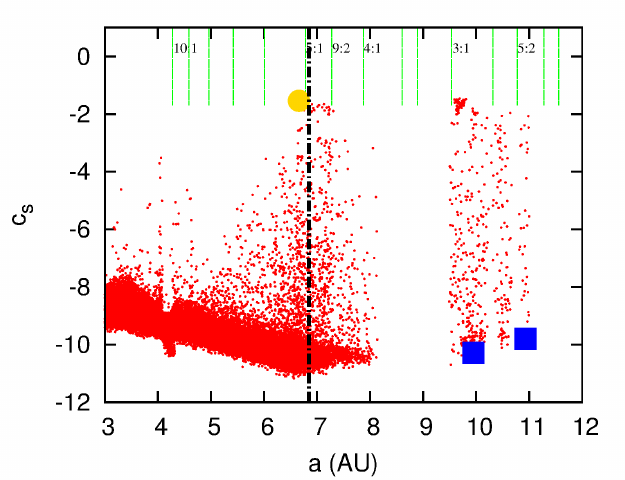}
  \caption{\label{a25_1pia_e00}
FMA analysis of a Jupiter--size single planet dynamics in a 
binary system with $a_B = 25$ AU and $e_B = 0$. The diffusion 
index $c_s$ is drawn vs. the semimajor axis of the planet.
Small values of $c_s$ means low diffusion, while large values 
connote chaotic orbits. 
The yellow filled circle is a case with a large 
diffusion index which becomes 
unstable after about 60 Myr. The two filled blue squares are
stable cases over 2 Gyr. The green dashed lines show the 
location of mean motion resonances between the planet and the 
companion star up to order 10. The black dash--dotted line marks 
the critical semimajor axis computed from the empirical 
formula of \cite{holman97}.}
\end{figure}

In Fig.\ref{a25_1pia_e00} we show the outcome of the FMA 
applied to a dynamical system composed of a
single planet orbiting the primary star of a binary system 
with $a_B = 25$ AU and 
$e_B = 0$. The red dots are all systems stable for at least 
10 Myr for which a value of diffusion speed $c_s$ has been computed. The 
black vertical line marks the empirical stability limit of \cite{holman97}
$a_c$. The green dashed vertical lines indicate the location of 
the major mean motion resonances with the binary companion. 
Close to $a_c$, instability begins to grow possibly due to the 
increasing strength of the resonances with the companion
star. The instability is
confirmed by the long term integration of a single case,  
marked by a yellow large filled circle in Fig.\ref{a25_1pia_e00},
where the planet is ejected from the system after about 60 Myr. 

However, 
stable systems are found more than 1 AU beyond $a_c$, 
out to about $a=8$ AU.
Thus,  
it appears that in this dynamical configuration $a_c$ underestimates 
the stability limit. 
In addition,  between 
9.5 to 11 AU a limited number of stable systems is also found
possibly trapped in mean motion resonances with the 
binary companion.
The numerical integration of two of these systems, marked by
large blue filled squares, shows that they are stable over 
4 Gyr. They show large regular oscillations in $a$ and $e$
due to the binary perturbations
and, since we plot the initial semimajor axis, their location 
in Fig.\ref{a25_1pia_e00} may be misleading. The average 
semimajor axis of these systems is shifted towards 
smaller values but they are still well beyond   
 9 AU.

\begin{figure}[hpt]
  \includegraphics[width=\hsize]{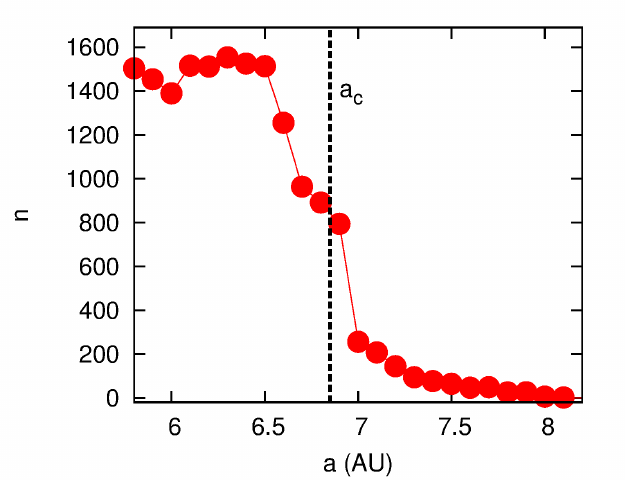}
  \caption{\label{a25_1pia_e00_histo}
Number of stable orbits ($c_s$ lower than -6) computed on equally spaced
discrete bins 0.1 AU wide. Close to $a_c$ there is a significant 
drop in number but stable orbits are present also beyond 
$a_c$ even if their number is small. To be found, these orbits 
require a fine sampling of the phase space which is possible
thanks to the use of the 
FMA analysis.
}
\end{figure}

What is the origin of the discrepancy between our results 
and the value $a_c$ computed from the formula of \cite{holman97}?
One possible cause is the difference in the dynamical model 
since we consider the full elliptical three--body problem 
with a Jupiter--size 
planet on a slightly inclined orbit while \cite{holman97} consider
massless particles on planar orbits. 
To test this hypothesis we repeated 
the calculations of Fig.\ref{a25_1pia_e00} setting the mass 
of the planet to $0$ in order to better compare with the outcome of 
\cite{holman97}. We find a higher instability in the 
full three--body model in particular close to the outer
stability limit, but the value of this limit is approximately
equal in the two models and larger than $a_c$. We then set 
the inclination of the massless particles to 0i, i.e. they 
all lie on the same plane of the binary but still there are bodies 
on stable orbits beyond $a_c$ as in Fig.\ref{a25_1pia_e00}.
A possible alternative explanation to this discrepancy is that 
we sample a much larger number of dynamical systems compared 
to \cite{holman97} leading to a finer sampling of the phase space.
If we compute the number of stable orbits ($c_s$ lower than 
-6) on discrete bins in semimajor axis  we can explore the 
fraction of stable orbits as a function of the planet semimajor axis. 
In Fig.\ref{a25_1pia_e00_histo} we plot the number of stable cases $n$ over 
 a range of semimajor axis centered on $a_c$.
$n$ is approximately stationary out to 6.5 AU, then it 
sharply declines till 6.8 AU which is the predicted value of 
$a_c$ according to\cite{holman97}. 
However, beyond $a_c$ there are additional stable orbits whose frequency is 
low compared to that within $a_c$. If the sampling of initial 
configurations is limited in number, the stable region  
beyond $a_c$ can be missed because not densely populated. 
This is possibly the reason of the difference between our 
results and those of \cite{holman97}.

\begin{figure}[hpt]
  \includegraphics[width=\hsize]{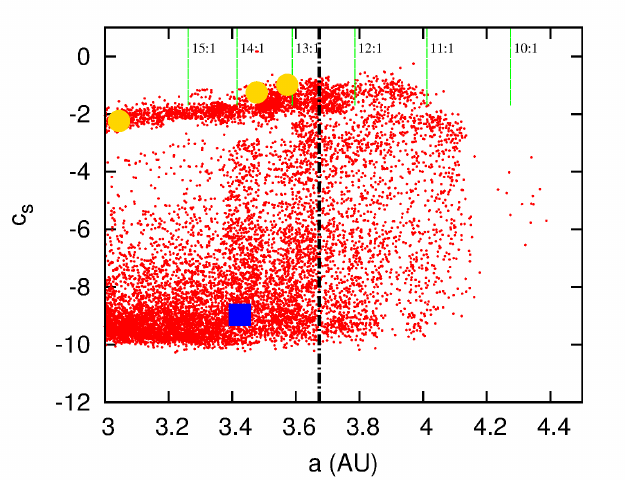}
  \caption{\label{a25_1pia_e04}
Same as Fig.\ref{a25_1pia_e00} but with 
$a_B = 25$ AU and $e_B = 0.4$. The larger eccentricity
of the binary system induces a higher level of chaoticity. }
\end{figure}

\begin{figure}[hpt]
  \includegraphics[width=\hsize]{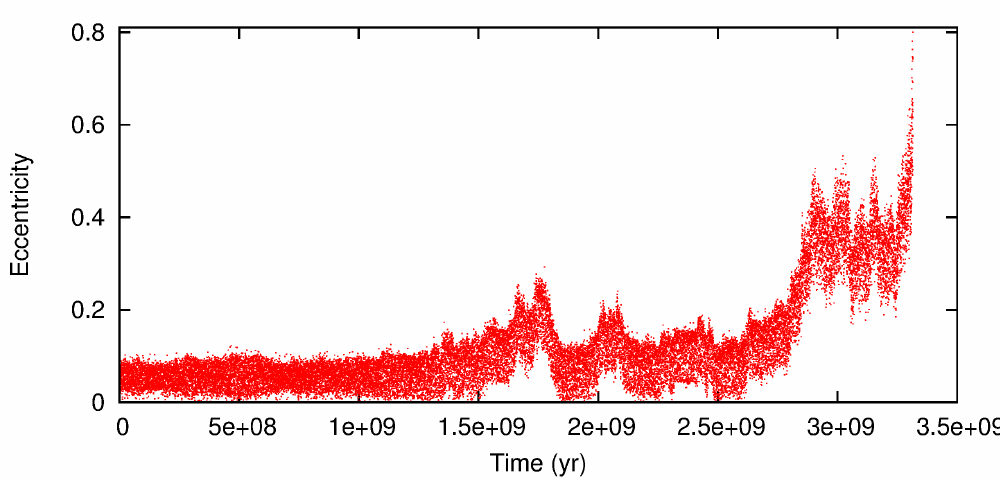}
  \caption{\label{single304_a25_e04}
Time evolution of the eccentricity of the system marked by a 
yellow circle in Fig.\ref{a25_1pia_e04} whose initial 
semimajor axis is close to 3.04 AU. The system is chaotic
and after about 3.3 Gyr the planet is ejected on a hyperbolic
trajectory. The origin of its instability is possibly due to 
a non--linear secular resonance.}
\end{figure}

To test the dependence of the stability analysis on the eccentricity of 
the binary we repeated the FMA for a model with $e_B = 0.4$,
the most frequent value encountered in binary systems according
to \cite{duma}. In Fig.\ref{a25_1pia_e04} the FMA outcome is 
illustrated in this case. Again, the limit of stability is 
wider compared to that predicted by \cite{holman97} 
since stable configurations are found when the semimajor 
axis of the planet is around 4.1 AU while $a_c = 3.67$ AU.
The most interesting feature is, however, the bimodal distribution
of the stability index $c_s$ as a function of the semimajor 
axis. As in Fig.\ref{a25_1pia_e00}, the yellow filled circles are 
unstable orbits while the blue filled squares are stable ones. 
The two outer unstable systems have the planet ejected on a 
hyperbolic trajectory in less than 60 Myr. The inner case with 
an initial semimajor axis around 3.04 AU takes more time and 
its evolution is shown in Fig.\ref{single304_a25_e04}. The 
orbit shows clear signs of chaotic behaviour from starting,
visible only by inspecting the orbit on a short timescale. Finally, the 
ejection of the planet from the system occurs after about 3.3 Gyr
possibly because the path to full instability is long. 

\begin{figure}[hpt]
  \includegraphics[width=\hsize]{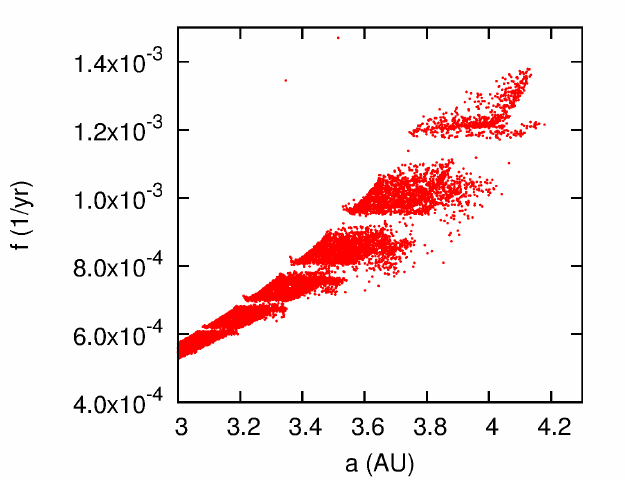}
  \caption{\label{a25_1pia_e04_freq}
Main frequency of the $h$ and $k$ variables of the planet,
as computed by the MFT method within the FMA analysis, vs. the 
semimajor axis. The empty stripes mark the location of 
non--linear secular resonances which appear only for 
high values of the binary eccentricity ($e_B = 0.4$).}
\end{figure}

To understand the reason of the coexistence of stable and 
unstable orbits for similar values of semimajor axis we have
to explore the frequency space. In Fig.\ref{a25_1pia_e04_freq} 
the main frequency of the planet orbit is shown as a function of 
the initial semimajor axis. There are empty stripes which 
possibly correspond to the families of unstable orbits identified 
by \cite{mich2004} as due to non--linear secular resonances 
which appear in their generalized numerical secular perturbation theory. 
These same resonances were retrieved at moderate eccentricities by 
the order 12 secular theory developed for two coplanar planets 
by \cite{libert2005}. 
There are significant differences between the configuration explored
by \cite{mich2004} and ours in particular concerning the mass ratio
between the two bodies of the system other than the primary star. 
\cite{mich2004} consider a mass ratio ranging from 
1 to 1/4 between two planets perturbing each other 
while in our case the ratio between the companion 
star and the planet is about $1 \times 10^3$. However, we may expect that
the origin of the instability stripes may be ascribed to similar
non--linear resonances which appear only when the 
companion star is on a highly eccentric orbit  ($e_B = 0.4$). 
In fact, for $e_B = 0$, there is no sign of similar unstable 
regions in the frequency vs. semimajor axis space and this is 
in agreement with the findings of \cite{mich2004} predicting 
the presence of secular resonances at moderate to large 
eccentricities. 

 Another way to estimate a stability limit for single planet orbits in binaries is 
  to use the Hill stability criterion described in \cite{marchal82}.
  We use its formulation as described in 
\cite{ines} and compute a radius of the stability sphere $R_s$  using their 
equations 3) and 4). Since the criterion is defined on the radial distance, 
for applying it to keplerian orbits it is necessary to adopt the following rule:
an orbit satisfies the Hill stability criterion if $a_{pl} (1 + e_{pl}) < R_s$, 
in other words the aphelion of the orbit must lie inside $R_s$. We applied this 
criterion to the two models with $a_B = 25$ AU and $e_B = 0$ and $e_B = 0.4 $, respectively. 
To compare our sample of orbits with $R_s$, we select from 
the whole sample those with 
$c_s < -6$,  as a conservative choice for stable orbits. For these we compute the 
maximum value of the aphelion distance over the integration timespan and compare 
it with $R_s$. For the case described in Fig.\ref{a25_1pia_e00}, i.e. $a_B=25$ AU and 
$e_B = 0$,  
we find that, all over the sample of stable orbits, the maximum aphelion distance 
is $9.6$ AU which is in good agreement with the prediction of eq. 4) in \cite{ines} for $\mu = 1$ 
which gives $R_s = 9.8$ AU. In the second case illustrated by 
Fig.\ref{a25_1pia_e04}, i.e. $a_B=25$ AU and      
$e_B = 0.4$ the maximum aphelion distance is $4.8$ AU against a value of 
$5.9$ AU predicted by the analytical equation.
The Hill criterion based on the definition of $R_s$ is assumed to be limited to 
'short--term' stability, as argued by \cite{ines}, however it is interesting that 
the discrepancy is larger for higher values of $e_B$.

\begin{figure}[hpt]
  \includegraphics[width=\hsize]{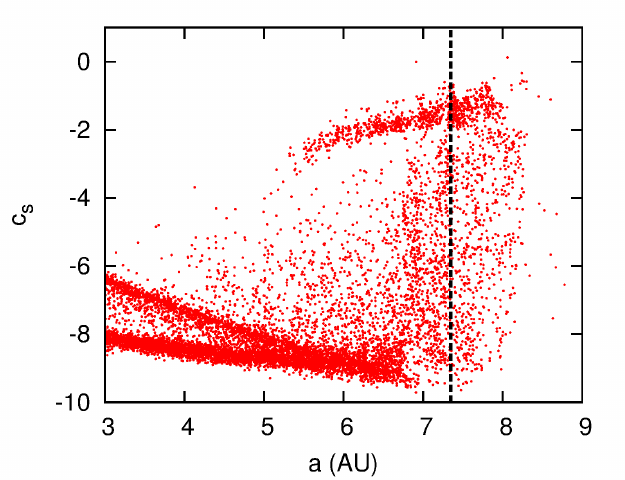}
  \caption{\label{a50_1pia_e04}
Same as Fig.\ref{a25_1pia_e04}  but with 
$a_B = 50$ AU and $e_B = 0.4$. The non--linear secular resonances 
are effective in inducing instability only close to the outer 
stability border and their effect is significantly reduced 
within $a \sim 5$ AU. Even in this case, the limit of stability 
extends beyond the empirical prediction of 
\cite{holman97}  (black dashed line).
}
\end{figure}

When the semimajor axis of the binary is increased to 
$a_B = 50$ AU, the stability portraits are very similar
to the $a_B = 25$ AU when the eccentricity of the binary
is set to 0. When $e_B = 0.4$, the non--linear resonances 
contribute to the instability also close to the outer 
border even if the major source of chaos in that region is the 
mean motion resonance overlap with the binary companion.
In Fig.\ref{a50_1pia_e04} the diffusion 
map of the planet orbits shows a bimodal distribution 
only close to the critical semimajor axis $a_c$ (black 
dashed line) but within 5 AU the effects of the 
non--linear resonances are strongly reduced. In the 
frequency space the same instability stripes, due to
these non--linear resonances, are observed as in 
Fig.\ref{a25_1pia_e04_freq}.

The test of the FMA method to explore the stability of 
single planetary systems in binary have shown to be 
effective and it has lead to new findings that improve
the analysis of \cite{holman97}. In particular, 
we observe that:

\begin{itemize} 

\item 
the value of $a_c$ computed with the 
empirical equation of \cite{holman97} slightly underestimates
the stability limit in the case of equal mass binary
systems. This is related to the need of a fine sampling of the 
phase space to outline the stability region. 

\item A limited number of stable systems is found well 
beyond $a_c$ and they are possible trapped into 
stable resonant 
orbits  with the binary companion. Their orbital 
elements show wide oscillations due to the companion 
perturbations.

\item When the eccentricity of the binary is set to 
$e_B = 0.4$, non--linear resonances cause instability 
for semimajor axis of the planet well within 
$a_c$ as shown in Fig.\ref{a25_1pia_e04}, 
Fig.\ref{a25_1pia_e04_freq} and 
Fig.\ref{a50_1pia_e04}. In highly eccentric binaries, 
the stability limit $a_c$ is then only indicative and 
unstable orbits can be found within it. 

\end{itemize}

We performed additional simulations
with $a_B = 25$ and $50$ AU and $e_B = 0.$ and $0.4$ 
but changing the mass ratio $\mu$ from $0.5$ to 
$\mu =0.33$ ($M_2 = 0.5 M_{odot}$)  and
$\mu = 0.17$ ($M_2 = 0.2 M_{odot}$).  
We compared the outer limit derived from the 
FMA with $a_c$ and found 
that the difference is always in between 10 to 20\%. From these 
results we conclude that on average $a_c$ is a good approximation 
but for a detailed study of a single planetary system 
a finer sampling of the phase space is needed since 
$a_c$ can be wrong by up to 20/

\section{Stability of two--planet systems}
\label{2pla}

We now apply the FMA method to systems made of two planets 
in orbit around the primary star of a binary systems. Our goal
is to test whether the stability properties outlined by
both $a_c$ and $\Delta$ are preserved or if it is necessary to
define limits for their applicability which depend on the binary configuration. 

\subsection{The circular case ($e_B = 0$).}

\begin{figure}[hpt]
  \includegraphics[width=\hsize]{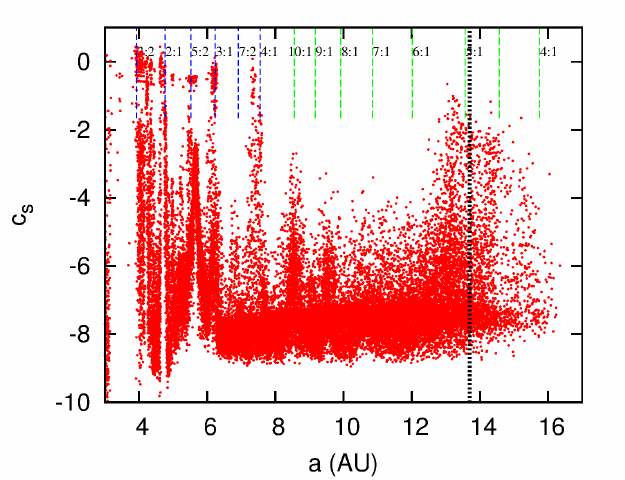}
  \caption{\label{fig_a50_2pia_e00}
Diffusion map of a two--planet system in a binary with $a_B = 50$ AU and 
$e_B = 0$. For $a_2 \leq 7$ AU mean motion resonances 
between the two planets  (blue dashed vertical lines) shape the stability portrait 
while beyond 
7 AU the resonances between the outer planet and the binary 
(green dashed vertical lines) 
determine the chaotic diffusion of the orbit. The black 
dotted line marks the location of $a_c$. 
}
\end{figure}

To explore the stability properties of 
two planet systems, we 
start with larger values of the binary separation 
$a_B$ since the semimajor axis of the inner planet is always set at 
3 AU. For small values of $a_B$, like $a_B = 25$, there
is little room for a stability analysis  since the sum of $a_1 + \Delta$
is larger than $a_c$. 
For this reason we begin with a binary separation $a_B = 50$ AU 
with the stars on a circular orbit ($e_B = 0$). 
In Fig.\ref{fig_a50_2pia_e00} we show the diffusion 
portrait of the system as a function of the semimajor axis of the 
outer planet $a_2$. Within about 7 AU, the mutual secular perturbations  of 
the two planets dominate the evolution of the system while, 
beyond that limit, the stability properties are determined 
by the perturbations of the companion star on the outer 
planet. As in the case with a single planet (Fig.\ref{a25_1pia_e00}),
chaotic orbits increase in frequency close to the 
critical stability limit $a_c$. However, as in the test case
with a single planet,  stable configurations
are found even for $a_2$ as far as $ \sim 16$ AU confirming that 
the analytical value of $a_c$ computed with the empirical 
formula of \cite{holman97} underestimates the real outer limit.
However, it appears that 
the presence of the inner 
planet does not significantly affect the stability properties
of the second planet close to 
$a_c$. 

\begin{figure}[hpt]
  \includegraphics[width=\hsize]{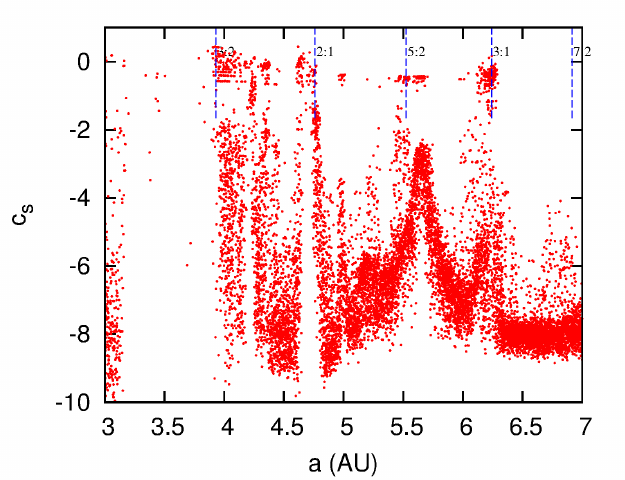}
  \includegraphics[width=\hsize]{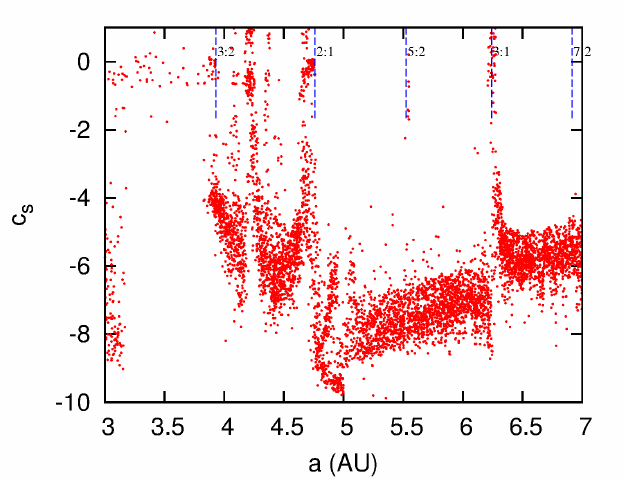}
  \caption{\label{fig_a50_2pia_e00_LARGE}
Enlarged view of the diffusion portrait shown in Fig.\ref{a25_1pia_e00}
limited to $a_2 \leq 7$ AU (upper panel).
The same diffusion map for a two planet
system around a single star is shown as a 
comparison in the lower panel. The vertical dashed 
lines mark the locations of mean motion resonances 
between the two planets. The dynamics in the binary 
system appears to be more chaotic in particular close to the 
inner stability border marked by the 3:2 mean motion resonance
between the planets and in correspondence to the 5:2 resonance. 
These effects are due to the gravitational perturbations of the 
binary companion. 
}
\end{figure}

An enlarged view of the diffusion map 
in Fig.\ref{fig_a50_2pia_e00}, limited to $a_2 \leq 7$ AU,
is shown in 
Fig.\ref{fig_a50_2pia_e00_LARGE} (upper panel). It is compared 
to a similar map without the binary companion (lower panel) 
to test the influence of the 
secondary star. 
This second plot is slightly different respect to that shown in 
\citep{mar20142pia} in particular close to the 3:2 resonance where stable 
orbits begin to appear. Even if the implications of the two 
plots for the 
long term stability of a two--planet system
are the same, in that shown in \citep{mar20142pia} the 
value of $c_s$ appear larger compared to similar configurations 
in Fig.\ref{fig_a50_2pia_e00_LARGE} 
lower panel. This difference is 
related to the use of 
shorter running windows  while applying the FMA ($5 \times 10^5$ yr
in this work, 2 Myr in \citep{mar20142pia}) since the binary
companion usually leads to faster secular frequencies compared to
systems made of two planets only. In addition,  our running windows shift by 
$1 \times 10^5$ yr while in \citep{mar20142pia} they were shifted of 
$1 \times 10^6$ yr possibly causing a larger spread of the computed 
frequency.
Two minor sources of differences are related to 
1) 
the numerical integrator used to compute the 
orbital evolution: in this study it is the 15th order 
RADAU integrator \cite{Everhart85} while in \citep{mar20142pia} 
the symplectic integrator SyMBA \citep{levdun94,levdun98}
was adopted (due to its Hamiltonian splitting structure, SyMBA cannot be
used in binary star systems)i and 2) to the selection of the FMFT algorithm 
instead of the MFT endorsed in \citep{mar20142pia}. 

The comparison between the upper and lower panel of 
Fig.\ref{fig_a50_2pia_e00_LARGE} shows that the 
the presence of the binary companion significantly 
increases the level of chaoticity close to the Hill stability limit
marked by the location of the 3:2 mean motion resonance between the planets.  
In the case with the binary companion (upper panel), the inner 
stability limit is populated by a significantly larger number 
of fast diffusing orbits respect to the same location in the 
phase space in the lower panel where the companion star has been removed 
from the system. 
Additional unstable 
features appear beyond 5 AU in the upper panel, more marked than 
in the case with a single star. The higher instability observed in 
the binary case is possibly due to the superposition of resonances. 
In addition to the mutual resonances between the planets, resonances 
with the companion star further perturb the system causing a 
more complex superposition effect. This is manifest not only 
close to the stability limit but also in correspondence to the
5:2 resonance between the planets which is markedly wider and 
chaotic in the binary case.  In both panels stable orbits for 
Trojan planets are observed close to 3 AU. The number of cases 
is too small to determine in detail the influence of the 
binary companion on their stability, 
however it looks like their existence is not jeopardized by
the presence of a secondary star. 
It is noteworthy that the diffusion speed of the orbits in 
the lower panel, where the binary companion is neglected, is 
increasing with $a_2$, the semimajor axis of the outer planet. 
This is not a dynamical effect but it is due to a degradation 
of the numerical  
precision in the determination of the secular frequency 
in each window. Such an effect, observed also in 
\cite{mar20142pia}, is due to the use of a fixed timespan for each 
running window while computing a frequency which is decreasing
for larger $a_2$. At the same time, the
forced eccentricity of each planet is smaller for larger values of $a_2$
leading to smaller values of the h, k variables, further increasing the
numerical error in the computation of the associated frequencies.
This systematic trend does not invalidate the 
analysis once its origin is known. In addition, in the case 
of two planets in binaries, both these effects are significantly reduced 
since the secular frequencies, due to the binary perturbations, 
are higher and the forced eccentricity grows for larger values of 
$a_2$.

A similar behaviour is observed also when $a_B = 100$ AU even if the level 
of chaoticity at the inner stability limit for the two planets is 
lower. In conclusion, the perturbations of the binary companion 
do not shift outwards the value of the minimum separation $\Delta$ 
of two planets 
for stability, but they significantly increase the frequency of chaotic 
orbits close to the border and even beyond at the major resonances
between the planets. 

\subsection{The eccentric case ($e_B = 0.4$).}

\begin{figure}[hpt]
  \includegraphics[width=\hsize]{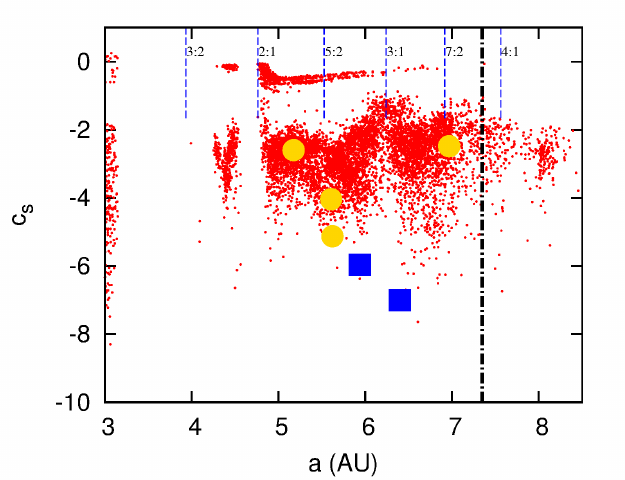}
  \caption{\label{fig_a50_2pia_e04}
Diffusion map of a two--planet system in a binary with $a_B = 50$ AU and 
$e_B = 0.4$. The phase space is mostly chaotic with few exceptions 
of low diffusion orbits. The yellow filled circles mark systems 
that become unstable followed by a period of close encounters 
between the planets. The blue squares are systems which are
stable over 10 Gyr. }
\end{figure}

When the eccentricity of the binary is increased to $e_B = 0.4$, 
in the case with $a_B = 50$ a 
two--planet system is unstable for almost all values 
of $a_2$ (of course assuming that $a_1 = 3$ AU). The stability portrait in 
Fig.\ref{fig_a50_2pia_e04} shows that only a limited number of 
systems have low diffusion rates compatible with long term survival. 
The yellow filled circles are planetary
systems which are chaotic ending with ejection of at least one planet
in less than 1 Gyr. 
They are  
located well within $a_c$ proving that the outer stability limit 
computed for systems with only one planet cannot be blindly 
applied to multiplanet systems. Only for very low values of 
$c_s$ a limited number of systems appear to be stable  
as those 
cases marked by a
blue filled square which display a quasi--periodic behaviour over 
5 Gyr. The mechanism responsible for the instability is 
again the superposition of mean motion and secular (linear and 
non--linear) resonances 
leading to overall instability with 
few exceptions.

\begin{figure}[hpt]
  \includegraphics[width=\hsize]{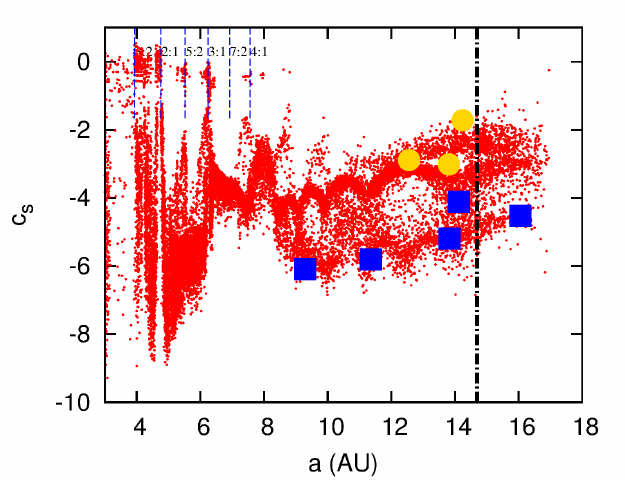}
  \caption{\label{fig_a100_2pia_e04_LARGE}
Diffusion portrait for a two--planet system in a binary with 
$a_B=100$ AU and $e_B=0.4$. The inner region, where the mutual planetary
resonances dominate, is more chaotic compared to the 
case with $a_B=50$ AU and $e_B=0$. Getting closer to the outer 
border, 
stable and unstable orbits coexist as in   
Fig.\ref{a25_1pia_e04} and Fig.\ref{a50_1pia_e04}.
}
\end{figure}

If we increase the binary semimajor axis to $a_B = 100$ AU 
(Fig.\ref{fig_a100_2pia_e04_LARGE}), the 
stability range extends out to $a_c$ and beyond. However, the 
region in proximity of the inner stability limit is more chaotic 
than in the $a_B = 50$ and $e_B = 0$ case (Fig.\ref{fig_a50_2pia_e00_LARGE})
with non--linear resonances contributing to instability while approaching  
$a_c$.

From the outcome of the FMA analysis, we can conclude that the 
eccentricity of the binary $e_B$ is more critical than $a_B$ 
in 
determining the chaotic nature of multiplanet systems. 
Moreover, the empirical formula of \cite{holman97}  is a 
good approximation 
when a single planet is orbiting the primary star
but  it may become very inaccurate for multiplanet systems
as illustrated in Fig.\ref{fig_a50_2pia_e04}. 

\section{Statistical analysis.}

\begin{figure}[hpt]
  \includegraphics[width=\hsize]{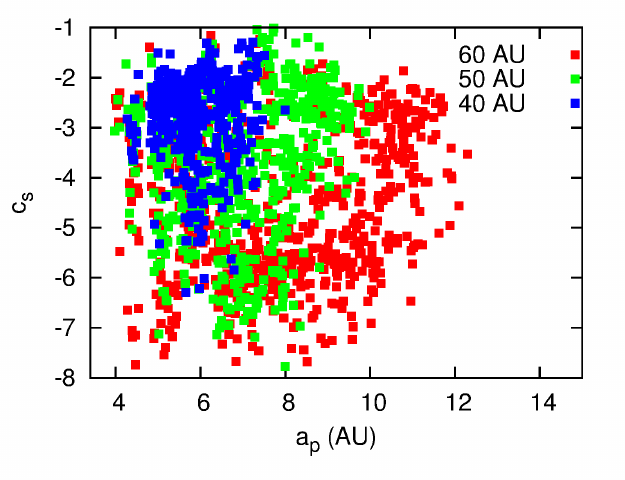}
  \caption{\label{stat0}
Diffusion portrait for a two--planet system in a binary with
$e_B = 0.3$, $\mu =1$, $a_1 = 3$ AU and for 3 different values of 
$a_B$, i.e. 40, 50 and 60 AU. A progressive increase in size of the stable region is
observed both inside, close to the inner planet, 
and outside. 
}
\end{figure}

By inspecting Fig.\ref{fig_a50_2pia_e04}, it appears that a system of two planets in binaries 
is not necessarily stable when both planets have semimajor axes smaller than $a_c$, the 
critical semimajor axis defined by \citep{holman97}. The combination of mean motion 
resonances between the two planets, between the planets and the binary companion and
non--linear secular resonances easily lead to chaotic evolution even within  $a_c$. 
The 
results shown in the previous section suggest that this may happen 
in particular for higher values of $e_B$.
To grant the presence of  a
stable region for two--planet systems, for a given value of 
$e_B$, the semimajor axis of the binary $a_B$ must be larger than a 
critical value. 
This is confirmed by 
Fig.\ref{stat0} where we compare three different cases with the same value of 
$e_B = 0.3$ but 3 distinct values of $a_B$, i.e. 40, 50 and 60 AU,
respectively. 
For $a_B = 30$ AU there are no stable systems while, by inspecting 
Fig.\ref{stat0}, 
it appears that for values of $a_B$ larger than 40 AU 
the stability of two planet systems is slowly restored within the 
critical value $a_c$ computed by Eq. 2). However, the filling of the 
region $a_2 < a_c$ with stable orbits is not instantaneous  but progressive and the fraction 
of stable orbits within $a_c$ increases with $a_B$.  This gradual 
stuffing of the region defined by $a_2 < a_c$ with stable orbits,
occurring for increasing $a_B$, is related to two distinct dynamical effects.
The first appears close to the inner planet, in between 4 and  6 AU, 
where there is a reduction of the 
chaoticity of the orbits marked by lower values of $c_s$
related to a weakening of superposition 
between mean motion resonances between the two planets and resonances between the 
planets and the binary.  
At the same time, the stability region stretches also out towards larger
values of $a_2$ for increasing values of $a_B$ because 
the overlap of mean motion and secular resonances with the companion become effective in 
generating chaotic behaviour farther out from the primary star. 

Taking advantage of this simple behaviour, we have tried to 
to derive a semi--empirical relationship
between $a_B$ and $e_B$ which can be predictive in terms of stability of two--planets
systems.  Our goal is to compute, for any given value of $e_B$,
a critical value of $a_B$, termed
$a_B^l$, 
below which any system of two planets is unstable. On the other hand, 
for $a_B > a_B^l$,  a significant fraction of stable bi--planet systems can 
be found with $a_2 < a_c$.  
Due to the large number of parameters playing a role in determining the 
stability properties of two planets in a binary we resorted to a statistical approach
for estimating $a_B^l$. 
First of all we sample different values of $e_B$ 
starting from $e_B = 0.1$ and incrementing it with a
step $\Delta e = 0.1$. For each value of $e_B$ we run the FMA analysis 
for 9 different values  of $a_B$ ranging from 10 to 100 AU at a pace of 
10 AU. Beyond 100 AU the secular evolution of the angles is too slow and an 
integration time longer than 10 Myr is needed for the FMA analysis.  We also consider 3 different 
values of the mass ratio, i.e. $\mu = 1, 0.5, 0.1$, and three different values of 
the semimajor axis of the inner planet $a_1 = 2, 3, 4$ in order to scale 
the value of $a_B^l$ even with these parameters.  

The number of simulations we ran is 
very large (more than $5 \times 10^5$) so we need a simple criterion to determine 
when a two--planet system 
has a sufficiently large stable zone 
for given values of $a_B$ and $e_B$.  
To 'measure' the size of the 
stable zone 
we resort to a method based on a simple Monte Carlo integration. 
We assume that the total volume in the phase space where stable two--planet 
systems can be found, for given $a_B$ and $e_B$, is limited by 
$a_c$, plus 20\% to account for the results presented in Sect. 3.
We then randomly sample 10000 systems for each $a_B, e_B$ 
for which $a_2 < a_c$, run the FMA analysis and finally count the number of systems 
having $c_s$ lower 
than -6 so that they are expected to be stable over a long timescale. 
The number of cases for which this last condition is fulfilled 
estimates the area of the stable zone.

\begin{figure}[hpt]
  \includegraphics[width=\hsize]{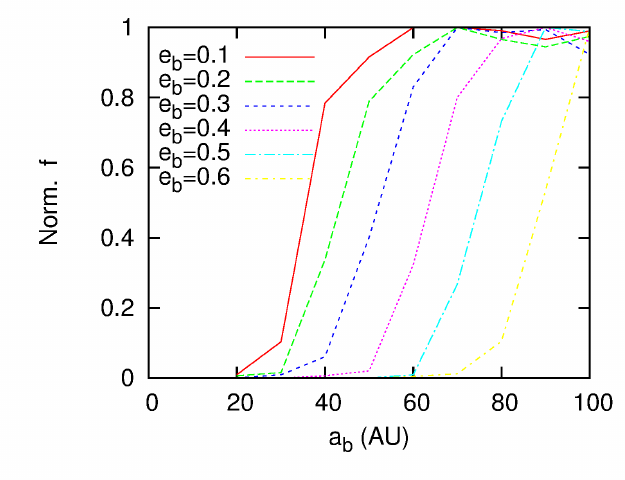}
  \caption{\label{stat1}
Normalized curves showing the number of stable orbits for 
different values of $a_B$ and $e_B$. The sudden step in the 
number of cases marked the onset of a significant 
stable zone for that binary parameters. The normalization is 
performed to the maximum number of stable cases. Since the 
models run at the same pace, we expect that $f$ measures the 
size of the stable region which saturates when the rate at which 
new stable cases are found is constant (i.e. the binary configuration
presents a large stable zone for two planets). 
 }
\end{figure}

In Fig.\ref{stat1} we show the number of stable cases as 
a function of both $a_B$ and $e_B$. For example, the red curve 
suggests that for $a_B < 20$ AU no stable two--planet 
systems can exist. The stable area begins to grow for $a_B =30$ AU, it has 
a sudden step marking a consistent widening of the stable zone 
for $a_B  =40$ AU,  and
it finally  
saturates at 60 AU where the stable region is only limited by the 
value of $a_c$. In this situation  the number of stable cases in the 10000 sample systems
is approximately the same even when larger values of $a_B$ are
considered. By inspecting Fig.\ref{stat1} we select the 
value of $a_B^l$ as the one for which the number of stable cases 
begins to grow.  
Of course, the resolution with which $a_B^l$ is computed  
is not very high since we use a step of 10 AU for $a_B$ and 
0.1 for $e_B$,  but we had to 
make compromises between the CPU load and resolution.
The data distribution shown in Fig.\ref{stat1} 
indicates that the number of stable orbits scales with the pericenter distance of 
the companion star, i.e. $q_B = a_B (1 - e_B)$ so we adopt a form of the 
fitting function reflecting this aspect. In addition, we perform an 
additional scaling on 
$\mu$ and $a_1$ obtaining the following fitting formula: 

\begin{equation}
a_b^l = {{10.00} \over {(1 - e_B)}} \cdot (1-0.54 (1-\mu^2)) \cdot a_1
\label{eq:fit}
\end{equation}

This equation is limited to the range of parameters
we have explored, i.e. $0 < e_b < 0.6$, $ 10 < a_B < 100$ AU, $0.1 < \mu < 1$,  and 
$2 < a_1 < 4$ AU. Adopting values of any of the three variables ($e_B, \mu, a_1$) 
far beyond these limits may lead to inaccurate values. For example, when 
$\mu$ reduces to 0.001, which is equivalent to assume that the binary companion
is a Jupiter--size planet, the value of $a_b^l$ should be about 8 AU, the 
value predicted by previous studies of the stability of three giant planets 
around single stars \citep{MW02}. The value that is obtained by the 
previous fit is instead 13.8 AU. However, we should define when a star is 
small enough to be termed a planet and vice versa. For this reason, we are
conservative when we claim that our fit is reliable within the 
limits given by the numerical simulations. 

\section{Discussion and conclusions}

Using the FMA, a fast tool to identify chaotic orbits, 
we have investigated the 
stability of planetary orbits in binary systems. 
As a test bench, we have first revisited the 
\cite{holman97} empirical formula which calculates a critical 
semimajor axis $a_c$ beyond which instability occurs 
for single planets in binaries.
We show that this formula underestimates 
the real stability limit by as much as 20\%, 
depending on the binary orbit. This is due to the 
low sampling rate used by \cite{holman97} which had to explore a 
wide range of parameters of the binary ($a_B$, $e_B$, and $\mu$) 
to perform a reliable fit based on these parameters.
As a consequence, for each individual set of $a_B, e_B, \mu$ they 
did not integrate a number of 
different systems large enough to fully explore the phase
space. We have focused 
on a limited number of binary parameters, but thanks to the FMA 
we have performed a very dense sampling finding that 
the stability limit in semimajor axis is larger than that given 
by the \cite{holman97}  
formula. The fraction of stable orbits decreases quickly 
beyond $a_c$ but they still represent a significant sample of 
stable configurations. In addition to these orbits, extending just
beyond $a_c$, there are additional dynamical systems which 
are trapped in resonance with the companion star, are stable over 
5 Gyr, and have semimajor axis well beyond $a_c$. 
When applied to the same systems, the Hill stability criterion applied
  to the binary companion and planet, 
on the other hand, predicts 
accurately the stability limit if the eccentricity of the 
binary is low but it becomes inaccurate for higher values of 
$e_B$ and it significantly overestimates the stability 
boundary.
The use of the FMA has also allowed us to discover unstable 
systems within $a_c$ when the orbit of the companion star is 
eccentric. These systems densely populate the 
region with  $a_p < a_c$ and their chaoticity is probably due to the presence of  
non--linear secular resonances \citep{mich2004,libert2005} 
induced by the companion star. 

When applied to two--planet systems, the FMA method shows that 
the separation between two planets leading to instability 
remains close to the Hill stability limit defined by \cite{glad93}.
However, when $a_B$ decreases and $e_B$ grows,
the perturbations of the companion star progressively make 
two--planet systems more chaotic. The inner border is populated 
by an increasing number of unstable systems while the outer border 
is destabilized by overlap of mean motion 
and secular (linear and non--linear) resonances.
The limiting case is 
when most planetary systems are chaotic independently of the 
separation between the two planets as in the case with 
$a_B = 50$ AU and  $e_B=0.4$ (the inner planet has semimajor axis $a_1=3$ AU
in all our models). In this configuration
mean motion resonances and secular resonances (linear and non--linear)
easily superimpose leading to chaos and instability. 
This implies that the value of $a_c$ cannot be carelessly used in 
multiplanet systems since the combined mutual planetary and 
binary perturbations may lead to chaotic behaviour well 
within $a_c$ in particular for eccentric binaries. 

To derive a semiempirical formula that allows us to compute, at least 
at a rough  approximation level, 
the value of the binary semimajor axis $a_B^l $ below which 
two planets cannot coexist around the central star,  
we have run a series of models with different 
values of $e_B, \mu, a_1$ with the goal of finding scaling relationships 
in these variables. 
Of course, owing to the huge amount of 
CPU load required to explore such a large parameter space, our 
formula is limited. However, it covers a range of parameters which are 
the most frequently encountered in binary systems, according to \cite {duma}. 
It can be used when planning observation strategies while looking for planets 
in binary systems.

\section*{ACKNOWLEDGMENTS}

We thank an anonymous referee for his/her useful comments and suggestions
which helped to improve the paper. 


\bibliographystyle{aa}
\bibliography{biblio}

\label{lastpage}

 \end{document}